\documentclass[10pt,aps,twocolumn,preprintnumbers,prd,noshowpacs,nofootinbib,noshowkeys,floatfix,superscriptaddress]{revtex4}
\usepackage[dvips]{graphics,graphicx}
\usepackage[colorlinks=true,linktocpage=true,linkcolor=blue,citecolor=blue]{hyperref}
\usepackage[usenames,dvipsnames]{color}
\usepackage{amsmath, amssymb,oldgerm}
\usepackage{multirow}
\usepackage{longtable}
\usepackage{color}
\usepackage[normalem]{ulem}  
\usepackage{braket}
\usepackage{slashed}
\usepackage[mathscr]{eucal}

\newcommand{\Tr}{\text{Tr}}
\newcommand{\n}{\nonumber}

\newcommand{\F}{\mathcal{F}}
\newcommand{\Pc}{\mathcal{P}}
\newcommand{\V}{\mathcal{V}}
\newcommand{\A}{\mathcal{A}}
\newcommand{\mC}{\mathfrak{C}}
\newcommand{\C}{\mathcal{C}}
\newcommand{\Sc}{\mathcal{S}}
\newcommand{\ms}{\mathfrak{s}}
\newcommand{\f}{\mathfrak{f}}

\newcommand{\beq}{\begin{equation}}
\newcommand{\eeq}{\end{equation}}
 \newcommand{\im}{\text{Im}\, }
 \newcommand{\re}{\text{Re}\, }

\newcommand{\inranglesm}{\rangle_\text{in}}

\newcommand{\outlanglesm}{{}_\text{out}\langle}

\begin{document}

\title{Generating spin polarization from vorticity through nonlocal collisions}

\author{Nora Weickgenannt}

\affiliation{Institute for Theoretical Physics, Goethe University,
Max-von-Laue-Str.\ 1, D-60438 Frankfurt am Main, Germany}

\author{Enrico Speranza}

\affiliation{Institute for Theoretical Physics, Goethe University,
Max-von-Laue-Str.\ 1, D-60438 Frankfurt am Main, Germany}

\author{Xin-li Sheng}

\affiliation{Peng Huanwu Center for Fundamental Theory and Department of Modern Physics,
University of Science and Technology of China, Hefei, Anhui 230026, China}

\author{Qun Wang}

\affiliation{Peng Huanwu Center for Fundamental Theory and Department of Modern Physics,
University of Science and Technology of China, Hefei, Anhui 230026, China}

\author{Dirk H.\ Rischke}

\affiliation{Institute for Theoretical Physics, Goethe University,
Max-von-Laue-Str.\ 1, D-60438 Frankfurt am Main, Germany}
\affiliation{Peng Huanwu Center for Fundamental Theory and Department of Modern Physics,
University of Science and Technology of China, Hefei, Anhui 230026, China}
\affiliation{Helmholtz Research Academy Hesse for FAIR, Campus Riedberg, Max-von-Laue-Str.\ 12, D-60438 Frankfurt am Main, Germany}

\begin{abstract}
We derive the collision term in the Boltzmann equation using the equation of motion for the
Wigner function of massive spin-1/2 particles. To next-to-lowest order in $\hbar$ it 
contains a nonlocal contribution, which is responsible
for the conversion of orbital into spin angular momentum. 
In a proper choice of pseudo-gauge
the antisymmetric part of the energy-momentum tensor arises solely from this 
nonlocal contribution.
We show that the collision term vanishes in global equilibrium and that 
the spin potential is then equal to the thermal vorticity.
In the nonrelativistic limit, the equations of motion for the energy-momentum and spin 
tensors reduce to the well-known form
for hydrodynamics for micropolar fluids.
\end{abstract}

\preprint{USTC-ICTS/PCFT-20-13}



\maketitle


\textit{Introduction} --
The study of polarization phenomena in heavy-ion collisions has attracted significant 
attention in the past~\cite{Liang:2004ph,Voloshin:2004ha,Betz:2007kg,Becattini:2007sr}.
Experimental studies show that the spin of hadrons emitted in noncentral collisions is
aligned with the direction of the global angular
momentum~\cite{STAR:2017ckg,Adam:2018ivw,Acharya:2019vpe}. The magnitude of the
global polarization of $\Lambda$ baryons can be very well described by models based on
relativistic hydrodynamics and
assuming local thermodynamic equilibrium of the spin degrees of
freedom~\cite{Becattini:2007sr,Becattini:2013vja,Becattini:2013fla,Becattini:2015ska,Becattini:2016gvu,Karpenko:2016jyx,Pang:2016igs,Xie:2017upb}. 
Particles get polarized through the
rotation of the medium in a way which resembles the Barnett effect~\cite{Barnett:1935}.
Unfortunately, the same models~\cite{Becattini:2017gcx,Becattini:2020ngo} are not able to
describe the experimentally measured longitudinal 
$\Lambda$ polarization~\cite{Adam:2019srw}.
This problem is currently the focus of intense work~\cite{Florkowski:2019qdp,Florkowski:2019voj,Zhang:2019xya,Becattini:2019ntv,Xia:2019fjf,Wu:2019eyi,Sun:2018bjl,Liu:2019krs}, 
however, a convincing solution does not yet exist.

In order to address this problem one needs to understand how the orbital angular 
momentum of the strongly interacting matter created in noncentral heavy-ion collisions 
is converted into the spin angular momentum of its constituents. In order to account for
the nontrivial dynamics of the spin degrees of freedom, it has been proposed to
introduce the rank-three spin tensor as additional dynamical variable, promoting relativistic
hydrodynamics to a theory of spin hydrodynamics~\cite{Florkowski:2017ruc,Florkowski:2017dyn,Florkowski:2018myy,Becattini:2018duy,Florkowski:2018fap,Hattori:2019lfp,Bhadury:2020puc} 
(for a different approach based on the Lagrangian formalism see Refs.~\cite{Montenegro:2017rbu,Montenegro:2017lvf,Montenegro:2018bcf,Montenegro:2020paq}).
Since spin is inherently a quantum feature, any rigorous derivation of 
spin hydrodynamics must be based on quantum field theory. 
However, in order to see how orbital angular momentum can be converted
into spin and vice versa on a microscopic level, it is advantageous to first derive a 
Boltzmann equation from quantum field theory, and then obtain hydrodynamics 
from the former, for instance by applying the method of moments~\cite{Denicol:2012cn}.

In the nonrelativistic case, this direction was 
pursued in a seminal paper~\cite{hess1966kinetic}, where
it was already pointed out that a local collision term in the Boltzmann equation 
will not be able to describe polarization through rotation. 
Only if one accounts for the nonlocality of the microscopic collision process,
orbital angular momentum can be converted into spin. A nonrelativistic nonlocal collision 
term was derived in Ref.\ \cite{hess1967verallgemeinerte} and for spinless particles, 
e.g., in Refs.~\cite{Morawetz:2017blm,Morawetz:2018shf,Morawetz:2000hr}.   
In Ref.~\cite{Zhang:2019xya}, a microscopic model based on collisions of partons 
as wave packets was proposed to link the spin polarization to the vorticity, 
but without considering spin equilibration.
A first attempt to systematically incorporate nonlocal collisions in a kinetic framework
based on quantum field theory was recently made in Ref.~\cite{Yang:2020hri}, 
however, without giving an explicit expression for the nonlocal collision term. 
The aim of this letter is to present the main steps of a systematic, explicit derivation of 
such a term, and
a discussion of its impact on spin hydrodynamics. For details of the calculation 
we refer to Ref.~\cite{Weickgenannt:2021cuo}.

Our notation and conventions are: $a\cdot b=a^\mu b_\mu$,
$a_{[\mu}b_{\nu]}\equiv a_\mu b_\nu-a_\nu b_\mu$, $g_{\mu \nu} = \mathrm{diag}(+,-,-,-)$,
$\epsilon^{0123} = - \epsilon_{0123} = 1$.


\textit{Quantum transport} --
The Wigner-function formalism provides a first-principle formulation of kinetic theory and 
is a very powerful tool for the description of anomalous transport in heavy-ion collisions
[see e.g.\ Refs.~\cite{Son:2012zy,Hidaka:2016yjf,Hidaka:2017auj,Huang:2018wdl,Gao:2018wmr,Yang:2018lew,Gao:2018jsi,Carignano:2018gqt}]. 
The Wigner function for spin-1/2 particles reads
\cite{DeGroot:1980dk,Heinz:1983nx,Vasak:1987um},
\begin{equation}
  W_{\alpha\beta}(x,p)=\int \frac{d^4y}{(2\pi\hbar)^4} e^{-\frac i\hbar p\cdot y}
  \left\langle :\bar{\psi}_\beta\left(x_1\right)\psi_\alpha\left(x_2\right):\right\rangle \; ,
\end{equation}
with $x_{1,2}=x \pm y/2$ and the spinor field $\psi (x)$.
Extending our previous work~\cite{Weickgenannt:2019dks} [see also 
Refs.~\cite{Fang:2016vpj,Florkowski:2018ahw,Gao:2019znl,Hattori:2019ahi,Wang:2019moi,Liu:2020flb}], 
which was valid in the free-streaming limit, we now include interactions and 
thus account for the effect of collisions.
Using the Dirac equation in the presence of a general interaction term
\begin{equation}
 (i\hbar\gamma\cdot \partial-m)\psi(x)=\hbar\rho(x)\; , \label{Dirac}
\end{equation}
where $\rho= - (1/ \hbar) \partial \mathcal{L}_I /\partial \bar{\psi} $, 
with $\mathcal{L}_I$ being the
interaction Lagrangian, the transport equation for the Wigner function 
reads \cite{DeGroot:1980dk},
\begin{equation}
 \left[ \gamma \cdot \left( p+i \,\frac{\hbar}{2} \partial \right) -m\right] W
 =\hbar\, \mathcal{C}[W]\; , \label{Wignerkin}
\end{equation}
where
\begin{equation}
 \C_{\alpha\beta}[W]\equiv \int \frac{d^4y}{(2\pi\hbar)^4} 
 e^{-\frac i\hbar p\cdot y}
 \left\langle :\rho_\alpha(x_2)\bar{\psi}_\beta(x_1):\right\rangle \;. 
\end{equation}
We decompose the Wigner function in terms of a basis of the generators of the 
Clifford algebra
\begin{equation}
W=\frac14\left(\F+i\gamma^5\Pc+\gamma \cdot \V+\gamma^5\gamma \cdot \A
+\frac12\sigma^{\mu\nu}\Sc_{\mu\nu}\right)\;, \label{dec}
\end{equation}
and substitute it into Eq.~\eqref{Wignerkin} to obtain equations of motion 
for the coefficient functions,
\begin{eqnarray}
\frac{\hbar}{2} \, \partial \cdot \A+m\Pc&=&-\hbar\, D_\Pc \;, \label{P}\\
p^{\mu} \F-\frac{\hbar}{2} \, \partial_\nu \Sc^{\nu\mu}-m\V^\mu&=&\hbar\, D_{\V}^{\mu} \;, 
\label{V}\\
\frac{\hbar}{2} \partial^{[\mu} \V^{\nu]}-\epsilon^{\mu\nu\alpha\beta}p_{\alpha}
\A_\beta-m\Sc^{\mu\nu}&=&\hbar\, D_{\Sc}^{\mu\nu} \;,\label{SSS}\\
 \hbar\,\partial \cdot \V&=&2\hbar\, C_\F \;, \label{Vkin}\\
p \cdot \A&=&\hbar\, C_\Pc \;, \label{orth}\\
 p^{[\mu} \V^{\nu]}+\frac{\hbar}{2} \, \epsilon^{\mu\nu\alpha\beta}\partial_\alpha
\A_\beta&=&-\hbar\, C_{\Sc}^{\mu\nu} \;, \label{Akin}
\end{eqnarray}
where we defined $D_i = \re \Tr\, (\tilde{\Gamma}_i \mathcal{C})$,
$C_i = \im \Tr\, (\tilde{\Gamma}_i \mathcal{C})$, $i = \F,\Pc,\V,\Sc$, 
$\tilde{\Gamma}_\F=1$,
$\tilde{\Gamma}_\Pc =-i \gamma_5$, $\tilde{\Gamma}_\V = \gamma^\mu$,
$\tilde{\Gamma}_\Sc= \sigma^{\mu \nu}$.

Following Refs.~\cite{Gao:2019znl,Weickgenannt:2019dks,Hattori:2019ahi} we employ
an expansion in powers of $\hbar$ for the functions 
$\F, \mathcal{P}, \V^\mu, \A^\mu, \Sc^{\mu\nu}$ and
the collision terms $D_i, C_i$ in Eqs.~\eqref{P}--\eqref{Akin}, e.g.\ for the scalar part 
\begin{equation}
	\F = \F^{(0)}+\hbar \F^{(1)} + \mathcal O (\hbar^2)\;.
	\end{equation}
Since gradients are always accompanied by factors of $\hbar$, 
this is effectively a gradient expansion.

In order to simplify the following discussion, we now make the assumption that all effects
arising from spin are at least of first order in $\hbar$.
Therefore, since $\A^\mu$ is the spin density in 
phase space~\cite{Weickgenannt:2019dks},
its zeroth-order contribution vanishes, $\A ^{(0)\mu}=0$, and consequently, 
from Eq.~\eqref{SSS}, $\Sc ^{(0)\mu\nu}=0$. 
From Eq.~\eqref{P} we also immediately conclude that $\Pc^{(0)} = 0$.
Thus, at zeroth order all pseudoscalar quantities vanish, which implies that also the
collision terms which carry pseudoscalar quantum numbers must vanish
at zeroth order, $D_\Pc ^{(0)}=C_\Pc^{(0)}=0$. 
Using Eqs.~\eqref{P} and \eqref{orth} this, in turn,
implies that
	\begin{equation}
	\Pc =\mathcal{O}(\hbar^2)\;, \;\;\; p\cdot \A=\mathcal{O}(\hbar^2)\,.   
	\label{spinorthcoll}
	\end{equation}
For the vector part, the only vector at our disposal at zeroth order is $p^\mu$, i.e.,
\begin{equation}
D_\V^\mu= p^\mu\delta V+\mathcal{O}(\hbar)\,,
\end{equation}
with a scalar function $\delta V$. Thus, from Eq.\ \eqref{V} we obtain
\begin{equation}
 \V^\mu=\frac{1}{m}p^\mu \bar{\F}+\mathcal{O}(\hbar^2)\,, \label{newV}
\end{equation}
where we defined $\bar{\F}\equiv \F-\hbar\delta V$.
From Eqs.\ \eqref{Vkin} and \eqref{Akin} we then derive
\begin{equation}
 	p\cdot\partial\bar{\F}=m\, C_F \;,  \;\;\;
  	p\cdot\partial\A^\mu=m\, C_A^\mu\; , \label{Fspintransport}
\end{equation}
with $C_F=2C_\F$ and 
$C_A^\mu\equiv -\frac{1}{m}\epsilon^{\mu\nu\alpha\beta}p_\nu C_{\Sc\alpha\beta}$.
Equations\ \eqref{spinorthcoll} and \eqref{Fspintransport} form a closed system
of equations for $\bar{\F}$ and $\A^\mu$.

The next step is to introduce spin as an additional variable in phase space
\cite{Zamanian:2010zz,Ekman:2017kxi,Ekman:2019vrv,Florkowski:2018fap,Bhadury:2020puc}.
We define the distribution function
 \beq
 \f(x,p,\ms)\equiv\frac12\left[\bar{\F}(x,p)-\ms\cdot\A(x,p)\right]\; . \label{spinproj}
 \eeq
We now employ the quasi-particle approximation, i.e.,
we assume that $\f$ is of the form
\begin{equation}
 \f(x,p,\ms)=m\delta(p^2-M^2) f(x,p,\ms)\;, \label{onshellsolution}
\end{equation}
where $f(x,p,\ms)$ is a function without singularity at 
$p^2=M^2 \equiv m^2+ \hbar\delta m^2$ and
$\delta m^2(x,p,\ms)$ is an interaction contribution to the
mass-shell condition for free particles
[where the $\ms$ dependence enters at $\mathcal{O}(\hbar)$]. We introduce the 
integration measure
\begin{equation}
 \int dS(p)  \equiv \frac{1}{\kappa(p)} \int d^4\ms\, \delta(\ms\cdot\ms+3)\delta(p\cdot \ms)\;,
\end{equation}
where $\kappa(p) \equiv \sqrt{3}\pi/\sqrt{p^2}$
is determined by requiring
\begin{equation}
 \bar{\F}=\int dS(p)\, \f(x,p,\ms)\;, \;\;\;
 \A^\mu= \int dS(p)\, \ms^\mu \f(x,p,\ms)\;.\label{A_int}
\end{equation}
The kinetic equation which we want to solve is
now given by combining
Eq.\ \eqref{spinproj} with Eq.\ \eqref{Fspintransport},
\begin{equation}\label{Boltzmannnn}
 p\cdot\partial \, \f(x,p,\ms)=m\, \mathfrak{C} \;,
\end{equation}
where $\mathfrak{C}\equiv \frac{1}{2}(C_F-\ms \cdot C_A)$.
In the next section, we will compute this kernel up to first order in $\hbar$.
This implies that we assume that $\f$ varies
slowly in space and time on the microscopic scale
corresponding to the interaction range. We will also restrict ourselves
to the contribution from particles; the extension to include antiparticles is straightforward.


\textit{Collisions} --
Up to first order in $\hbar$,
$\mathfrak{C}$ has the following structure,
\begin{equation} \label{fullcollterm}
	\mathfrak{C} =  \mathfrak{C}^{(0)}_{{l}}+\hbar \left\{ \mathfrak{C}^{(1)}_{{l}}
	+ \mathfrak{C}^{(1)}_{{nl}} \right\}
	 \equiv  \mathfrak{C}_{{l}} +\hbar \mathfrak{C}^{(1)}_{{nl}}\;.
\end{equation}
Here, local and nonlocal contributions are denoted by subscripts ${l}$ and ${nl}$,
respectively. The zeroth-order contribution is purely local~\cite{Li:2019qkf}, 
while the first-order contribution
has both local and nonlocal parts, the latter arising from gradients.

In order to explicitly calculate the collision term we follow Ref.~\cite{DeGroot:1980dk}.
This calculation is based on an expansion in particle-scattering states and makes the 
following assumptions: (i) low-density approximation, i.e., the interacting Wigner function 
in the collision term is identified with the free-streaming Wigner function 
(containing also contributions up to order $\hbar$) and only binary scattering 
processes of the form $(p_1,\ms_1),(p_2,\ms_2)\rightarrow (p,\ms),(p^\prime,\ms^\prime)$ 
are considered, and (ii) initial correlations are neglected.

The details of the calculation {are shown in Ref.\ \cite{Weickgenannt:2021cuo}.}
Here we only report the most important intermediate steps.
First, it can be shown that the off-shell terms on both sides of 
Eq.\ \eqref{Boltzmannnn} cancel
and we are left with the on-shell Boltzmann equation
\begin{equation}
\label{bbeqeq}
 \delta(p^2-m^2)\,p\cdot\partial f(x,p,\ms)= \delta(p^2-m^2)\,
 \mathfrak{C}_\text{on-shell}[f]\; . 
\end{equation}
where $\mathfrak{C}_\text{on-shell}[f]$ is the on-shell contribution of 
Eq.\ \eqref{fullcollterm}.
{In this way} we obtain the local part of the on-shell collision term
\begin{equation} \label{localcoll_1}
\mathfrak{C}_{\text{on-shell},l}[f]\equiv  \mathfrak{C}_{\mathrm{p+s}}[f] 
+  \mathfrak{C}_{\mathrm{s}}[f]\;,
\end{equation}
where 
\begin{subequations} \label{looocalcoll}
\begin{eqnarray}
\lefteqn{ \mathfrak{C}_{\mathrm{p+s}}[f]  \equiv  
\int d\Gamma_1\, d\Gamma_2\, d\Gamma^\prime \,
 dS_1^\prime(p)  \, \mathcal{W}} \\
 &\times& \big[  f(x,p_1,\ms_1)\big. f(x,p_2,\ms_2)\big.
 - f(x,p,\ms_1^\prime)f(x,p^\prime,\ms^\prime)\big] \n
 \end{eqnarray}
describes momentum- and spin-exchange interactions,
while 
\begin{equation}
\mathfrak{C}_{\mathrm{s}}[f] \equiv  \int  d \Gamma_2\, dS_1(p)\,
\mathfrak{W}\, f(x,p,\ms_1)f(x,p_2,\ms_2)
\end{equation}
\end{subequations}
describes purely spin exchange without momentum exchange.
Here we defined the phase-space measure
$\int d\Gamma \equiv \int d^4p\, \delta(p^2 - m^2) \int dS(p)$
as well as 
\begin{eqnarray}
  \lefteqn{\mathcal{W}\equiv \frac{1}{32} \sum_{s,r,s_1^{\prime}} \left[
  h_{s s_1^\prime} (p,\ms_1^\prime) h_{s_1^\prime r}(p,\ms) +
  h_{s s_1^\prime}(p,\ms) h_{s_1^\prime r} (p,\ms_1^\prime)\right] }\n \\
  && \times \sum_{s',r',s_1,s_2,r_1,r_2} h_{s^\prime r^\prime}(p^\prime, \ms^\prime) \,
  h_{s_1 r_1}(p_1, \ms_1)
  \, h_{s_2 r_2}(p_2, \ms_2) \n \\
    &&\times \langle{p,p^\prime;r,r^\prime|t|p_1,p_2;s_1,s_2}\rangle
  \langle{p_1,p_2;r_1,r_2|t^\dagger|p,p^\prime;s,s^\prime}\rangle\n\\
  && \times \delta^4(p+p^\prime-p_1-p_2)\;,\label{local_col_GLW}
\end{eqnarray}
and
\begin{eqnarray}
 \mathfrak{W}&\equiv&\hbar \frac{\pi}{4m} \sum_{s_1,s_2,r,r_2}
 \epsilon_{\mu\nu\alpha\beta} \ms^\mu \ms_1^{\nu} p^\alpha
 n_{s_1r}^{\beta}(p)\, h_{s_2r_2} (p_2, \ms_2)\n\\
 &&\times\langle{p,p_2;r,r_2|t+t^\dagger|p,p_2;s_1,s_2}\rangle \;,\label{mathw}
\end{eqnarray}
where
\begin{equation}
h_{sr}(p, \ms) \equiv \delta_{sr} + \ms \cdot n_{sr}(p)\;,\; \;
n^\mu_{sr}(p)\equiv \frac{1}{2m}\bar{u}_s(p)\gamma^5\gamma^\mu u_r(p)\;,
\end{equation}
with the standard particle spinors $u_r(p)$, $\bar{u}_s(p)$.
Moreover, 
\begin{align}
&\langle{p,p^\prime;r,r^\prime|t|p_1,p_2;s_1,s_2}\rangle\equiv\n\\
&-\sqrt{\frac{(2\pi\hbar)^{7}}{2}}\bar{u}_r(p) 
\outlanglesm p^\prime;r^\prime|\rho(0)| p_1,p_2;s_1,s_2\inranglesm
 \label{turho}
\end{align}
denotes the
amplitude for the scattering of two particles with momenta $p_1, p_2$ and 
spin projections $s_1,s_2$
into two particles with momenta $p, p'$ and spin projections $r, r'$.

If the distribution functions do not depend on the spin variables, i.e., $f(x,p,\ms)
\equiv f(x,p)$, we recover the familiar Boltzmann collision term, where the 
spin averaging and summation is done directly in the cross section~\cite{DeGroot:1980dk}.
However, if the distribution functions depend on spin, the two terms on the
right-hand side of Eq.~\eqref{localcoll_1} require further discussion.
Considering $\mathfrak{C}_{\mathrm{p+s}}[f]$, the term 
$\sim f(x,p_1,\ms_1)f(x,p_2,\ms_2)$
has the form of a gain term for particles with $(p,\ms)$, while
$\sim f(x,p,\ms_1^\prime)f(x,p^\prime,\ms^\prime)$ 
does not have an obvious interpretation as a loss
term, because the spin variable is $\ms_1^\prime$, not $\ms$.

However, we can modify the definition of distribution function and collision term
such that the standard interpretation of gain and loss terms in the latter is recovered.
To this end, we first note that the physically relevant quantities are obtained after 
integrating over the phase-space spin variable, see e.g.\ Eq.~\eqref{A_int}. 
Therefore, if we can replace
the distribution function $f (x,p,\ms)$ by another distribution function $\tilde{f}(x,p,\ms)$ and
similarly the collision term
$\mC[f]$ by $\tilde{\mC}[\tilde{f}]$, where
\begin{subequations} \label{wep}
\begin{eqnarray}
 p \cdot \partial \, \tilde{f}(x,p,\ms)&=&\tilde{\mC}[\tilde{f}]\;, \label{prop1}\\
 \int dS(p)\, b\, \tilde{Q}(x,p,\ms)&=& \int dS(p)\, b\, Q(x,p,\ms) \label{prop2}
\end{eqnarray}
\end{subequations}
is fulfilled for $Q\in \{f,\mC[f]\}$, $\tilde{Q}\in \{\tilde{f},\tilde{\mC}[\tilde{f}]\}$,
$b\in \{1,\ms^\mu\}$, then the physically relevant quantities will not be changed.
Equations~\eqref{wep} constitute a kind of ``weak equivalence
principle'', stating that $f$ and $\tilde{f}$ formally obey the same equation of motion and
give identical results when integrating over the spin variable.

One can show that the choice $\tilde{f} \equiv f$ and
\begin{eqnarray}
 \lefteqn{\tilde{\mC}_{\mathrm{p+s}}[f] \equiv  
 \int d\Gamma_1\, d\Gamma_2\, d\Gamma^\prime\,
 \widetilde{\mathcal{W}}} \n \\
 & \times & \!\!\!\! \big[ f(x,p_1,\ms_1)f(x,p_2,\ms_2) 
 -f(x,p,\ms)f(x,p^\prime,\ms^\prime)\big]\,,
 \label{localcollafter}
\end{eqnarray}
with
\begin{eqnarray}
 \lefteqn{ \widetilde{\mathcal{W}}\equiv\delta^4(p+p^\prime-p_1-p_2)\,\frac{1}{8} \sum_{s,r}
  h_{s r} (p,\ms) } \label{local_col_GLW_after} \\
  && \times \sum_{s',r',s_1,s_2,r_1,r_2} h_{s^\prime r^\prime}(p^\prime, \ms^\prime) \,
  h_{s_1 r_1}(p_1, \ms_1)
  \, h_{s_2 r_2}(p_2, \ms_2) \n \\
    &&\times \langle{p,p^\prime;r,r^\prime|t|p_1,p_2;s_1,s_2}\rangle
  \langle{p_1,p_2;r_1,r_2|t^\dagger|p,p^\prime;s,s^\prime}\rangle \;,\n
\end{eqnarray}
satisfies the weak equivalence principle \eqref{wep} up to $\mathcal{O}(\hbar)$.

Let us now focus on $\mC_{\mathrm{s}}[f]$.
This corresponds to collisions where the momentum of each particle
is conserved, but the spin can change: 
$(p,\ms_1),(p_2,\ms_2)\rightarrow (p,\ms),(p_2,\ms^\prime)$
\cite{DeGroot:1980dk}.
Here, the distribution functions $f(x,p,\cdot)$ and $f(x,p^\prime,\cdot)$ 
describe the particles
before \textit{and} after the collision, which means that they contribute to 
\textit{both} the gain and the
loss term. We see from Eq.~\eqref{mathw} that the interchange of $\ms^\mu$ 
and $\ms^\nu_1$ flips the
sign of $\mathfrak{W}$. This means that a net gain of particles with ($p,\ms$) corresponds
to a net loss of particles with $(p,\ms_1)$.

Turning now to the nonlocal collision term, let us first give an intuitive argument. 
If we assume that particles scatter with a finite impact parameter, 
the distribution functions entering Eq.\ \eqref{looocalcoll}
have to be evaluated at different space-time points. Hence, the simplest extension 
of the full collision term \eqref{fullcollterm}, and modified using the weak 
equivalence principle \eqref{wep}, should have the form
\begin{align}
 &\tilde{\mC}_{\text{on-shell}}[f]  =\int d\Gamma_1 d\Gamma_2 d\Gamma^\prime\,    
 \widetilde{\mathcal{W}}\,  [f(x+\Delta_1,p_1,\ms_1)\label{finalcollisionterm}\\
 & \times f(x+\Delta_2,p_2,\ms_2)-f(x+\Delta,p,\ms)f(x+\Delta^\prime,p^\prime,\ms^\prime)]
 \n\\
  &+\int  d\Gamma_2 \, dS_1(p)\,\mathfrak{W}
  f(x+\Delta_1,p,\ms_1)f(x+\Delta_2,p_2,\ms_2)\;, \n
\end{align}
where the position shifts $\Delta, \, \Delta',\, \Delta_1, \, \Delta_2$
are of first order in $\hbar$. Note that, in the collision term where only spin is exchanged,
$p_1 = p,\, p^{\prime} = p_2$.
We {show in Ref.\ \cite{Weickgenannt:2021cuo}} 
that Eq.~\eqref{finalcollisionterm} can indeed be
derived by an explicit calculation. The only additional assumption that we need to make 
is that the scattering amplitude is constant over scales of order $\Delta$.
This is consistent with the low-density approximation, see e.g.\ Ref.\ \cite{Abrikosov1975}.
The position shifts are functions of momentum and spin given by
\begin{equation}
\Delta^\mu\equiv -\frac{\hbar}{2m(p\cdot\hat{t}+m)}\, 
\epsilon^{\mu\nu\alpha\beta}p_\nu \hat{t}_\alpha \ms_{\beta}\;,
\end{equation} 
where $\hat{t}^\mu=(1,\boldsymbol{0})$ is the time-like unit vector in the frame where
$p^\mu$ is measured, for details see Ref.\ \cite{Weickgenannt:2021cuo}. 

\textit{Equilibrium} -- We will now consider the conditions necessary to reach equilibrium.
For the sake of simplicity, we consider uncharged particles, 
implying zero chemical potential.
The standard form of the local equilibrium distribution function 
is~\cite{Becattini:2013fla,Florkowski:2017ruc,Florkowski:2018fap}
\begin{equation}\label{f_eq}
 f_{eq}(x,p,\ms)=\frac{1}{(2\pi\hbar)^3}\exp\left[-\beta (x)\cdot p
 +\frac\hbar4 \Omega_{\mu\nu}(x)\Sigma_\ms^{\mu\nu}\right]\;,
\end{equation}
where $\beta^\mu=u^\mu/T$ ($u^\mu$ is the fluid velocity and $T$ the temperature, 
respectively) and
$\Omega^{\mu\nu}$ is the spin potential~\cite{Florkowski:2017ruc,Becattini:2018duy}.
Note that the exponent of the distribution
function has to be a linear combination of the conserved quantities momentum 
and total angular momentum. Here, we absorbed the orbital part of the 
angular momentum into the definition of
$\beta^\mu(x)$~\cite{Becattini:2013fla} and we defined the dipole-moment tensor
\begin{equation}
 \Sigma_{\ms}^{\mu\nu}\equiv -\frac1m \epsilon^{\mu\nu\alpha\beta}p_\alpha \ms_\beta\;.
\end{equation}
Inserting Eq.~\eqref{f_eq} into Eq.\ \eqref{finalcollisionterm} and expanding to first
order in $\hbar$ we obtain
\begin{align}
 &\tilde{\mathfrak{C}}_{\text{on-shell}}[f_{eq}]= - 
 \int d\Gamma^\prime d\Gamma_1 d\Gamma_2  \,
 \widetilde{\mathcal{W}} \,  e^{-\beta\cdot(p_1+p_2)}\n\\
 &\times  \left[\partial_\mu\beta_\nu \frac{}{}
 \left(\Delta_1^\mu p_1^\nu+\Delta_2^\mu p_2^\nu-\Delta^\mu p^\nu-\Delta^{\prime\mu}
 p^{\prime\nu} \right)
 \right.\n\\
 &\hspace*{0.4cm} \left. - \frac\hbar4\Omega_{\mu\nu}\left(\Sigma_{\ms_1}^{\mu\nu}
 +\Sigma_{\ms_2}^{\mu\nu}
 -\Sigma_{\ms}^{\mu\nu}-\Sigma_{\ms^\prime}^{\mu\nu}\right) \right]\n\\
 &-\int  \! d\Gamma_2\,  dS_1(p) dS^\prime(p_2)\, \mathfrak{W} \, 
 e^{-\beta\cdot(p+p_2)}\n \\
 & \times \left\{\partial_\mu \beta_\nu \left[(\Delta_1^\mu - \Delta^\mu) p^\nu 
 + (\Delta_2^{\mu} - \Delta^{\prime \mu}) p_2^{\nu} \right] \frac{}{} \right.\n\\
   &\hspace*{0.4cm} - \left.\frac\hbar4 \Omega_{\mu\nu}(\Sigma_{\ms_1}^{\mu\nu}
   +\Sigma_{\ms_2}^{\mu\nu}-   \Sigma_{\ms}^{\mu\nu}-\Sigma_{\ms^\prime}^{\mu\nu})
   \right\}\;,
    \label{colleqq}
\end{align}
where the zeroth-order contribution to the collision term was omitted, as it
vanishes for the distribution function \eqref{f_eq}.
Since $L^{\mu\nu}=\Delta^{[\mu}p^{\nu]}$ is the orbital angular momentum tensor of the
particle with $(p,\ms)$, the parentheses
in the second and fifth line contain the balance of orbital angular momentum 
in the respective collision.

Introducing the total angular momentum
$J^{\mu\nu}=L^{\mu\nu}+\frac\hbar2\Sigma_\ms^{\mu\nu}$ of the particle
and assuming that this is conserved in a collision,
$J^{\mu\nu}+J^{\prime\mu\nu}=J_1^{\mu\nu}+J_2^{\mu\nu}$,
the collision term vanishes for any $\widetilde{\mathcal{W}}$, $\mathfrak{W}$ if:
\begin{align}
\partial_{\mu}\beta_{\nu}+\partial_{\nu}\beta_{\mu}&=0\;, \\
\Omega_{\mu\nu}=\varpi_{\mu\nu}&\equiv-\frac12 \partial_{[\mu}\beta_{\nu]} 
= \mathrm{const.}\;,
\label{cond2}
\end{align}
which corresponds to the conditions for global (and not just local) equilibrium.
Our calculation derives condition~\eqref{cond2} for the first time in
a kinetic-theory approach, confirming a known result from statistical
quantum field theory~\cite{Becattini:2012tc}.
In previous works condition \eqref{cond2} was found in the 
massless case \cite{Chen:2015gta}, but for massive particles only in the presence of an
electromagnetic field~\cite{Weickgenannt:2019dks,Liu:2020flb}. 

\textit{Spin hydrodynamics from kinetic theory} --
The equation of motion for the spin tensor $S^{\lambda, \mu \nu}$
is derived from the conservation of the total angular
momentum tensor $J^{\lambda, \mu \nu} \equiv
x^\mu T^{\lambda \nu} - x^{\nu}T^{\lambda \mu} + \hbar S^{\lambda, \mu \nu}$~\cite{Florkowski:2017ruc,Florkowski:2017dyn,Florkowski:2018myy,Becattini:2018duy,Florkowski:2018fap,Hattori:2019lfp,Bhadury:2020puc}.
Among all possible pseudo-gauges~\cite{Hehl:1976vr,Leader:2013jra,Becattini:2018duy},
we use the energy-momentum and spin tensors proposed by Hilgevoord and Wouthuysen 
(HW)~\cite{HILGEVOORD19631,hilgevoord1965covariant,DeGroot:1980dk} which, unlike 
the canonical currents, were shown to give a covariant description for the 
spin of free fields. Defining
$dP\equiv d^4p\, \delta(p^2-M^2)$ one obtains
\begin{eqnarray} \label{KleinGordontensors}
 T^{\mu\nu}_{\mathrm{HW}}&=&\!\!\!\int dP\, dS(p)\,  p^\mu p^\nu f(x,p,\ms)  
 +\mathcal{O}(\hbar^2)\;,\\
 S^{\lambda,\mu\nu}_{\mathrm{HW}}
  &=& \!\!\! \int dP\, dS(p) p^\lambda\left(\frac{1}{2} \Sigma_\ms^{\mu\nu}-\frac{\hbar}{4m^2}
  p^{[\mu}\partial^{\nu]}\right) f(x,p,\ms)\n \\
  &  & + \mathcal{O}(\hbar^2)\;. \label{SpinHW}
 \end{eqnarray}
The equations of motion for these tensors can be written with the help of the
Boltzmann equation \eqref{bbeqeq} and the weak equivalence principle \eqref{wep} as
\begin{align} \label{Ta}
 \partial_\mu T^{\mu\nu}_{\mathrm{HW}} &  =  \int d^4p\, dS(p)\, p^\nu\, {\mC} = 0\;,\\
  \hbar\, \partial_\lambda S_{\mathrm{HW}}^{\lambda,\mu\nu}
 & = \int d^4p\, dS(p) \frac\hbar2 \Sigma_\ms^{\mu\nu}\, {\mC} 
 = T_{\mathrm{HW}}^{[\nu\mu]}\;, \label{total_cons}
\end{align}
where energy-momentum conservation in a binary collision makes the right-hand side of 
Eq.\ \eqref{Ta} vanish. On the other hand, the right-hand side of Eq.\ \eqref{total_cons} 
is derived from
the conservation of total angular momentum and shows that spin is in general not
conserved in collisions. This is described by the antisymmetric part of the 
energy-momentum
tensor, which contains terms of at least second order in $\hbar$. 
Moreover, in the HW formulation,
$T_{\mathrm{HW}}^{[\nu\mu]}$ is nonzero only in the presence of nonlocal collisions. In
local collisions the orbital angular momentum vanishes and the dipole-moment tensor itself
is a collision invariant~\cite{Florkowski:2018fap}, 
which makes the right-hand side of Eq.\ \eqref{total_cons} vanish.
In general, however, this happens only in global equilibrium. Away from global
equilibrium, the collision term is nonzero, and thus there is always dissipative dynamics.
In this sense, ``ideal spin hydrodynamics'' exists only if one
neglects the nonlocality of a microscopic collision.

\textit{Nonrelativistic limit} --  
In the nonrelativistic limit, 
$p^\mu\rightarrow m(1,\mathbf{v})$ with the particle three-velocity $\mathbf{v}$,
which implies $\Sigma_\ms^{ij}\rightarrow \epsilon^{ijk}\ms^k$. 
With this, the spatial components of
the last equality in Eq.\  \eqref{total_cons} read
\begin{equation}\label{Ta_nonrel}
 T_{\mathrm{HW}}^{[ji]}
 = m\epsilon^{ijk}\partial^0 \Big\langle{\frac\hbar2\ms^k}\Big\rangle + m\epsilon^{ijk}
 \partial^l \Big\langle{v^l \frac\hbar2\ms^k}\Big\rangle\;,
\end{equation}
where we used the Boltzmann equation \eqref{bbeqeq} to replace the collision term
$\tilde{\mC}[f]$ and introduced the notation
$\langle ...\rangle\equiv (m^2/2\pi\sqrt{3})\int d^3v\, d^3 \ms\, 
\delta (\boldsymbol{\ms}^2-3)\, (...)
f$. The above result agrees with Eq.\ (12.11) {of} Ref.\
\cite{hess1966kinetic} (up to a constant due to a different normalization).
We also compare to the results obtained for micropolar fluids.
The equation of interest here is Eq.\ (2.2.9) of Ref.~\cite{Lukaszewicz1999}, 
which in the absence of external fields reads
\begin{equation} \label{micro}
 \rho \left( \partial^0+u^j \partial^j\right)  \ell^i =  \partial^j C^{ji}+\epsilon^{ijk}T^{jk}\;,
\end{equation}
where $\rho$ is the mass density, $\ell^i$ is the internal angular momentum,
$C^{ji}$ is the so-called couple-stress tensor, 
and $T^{jk}$ is the (conventional) stress tensor.
Using the continuity equation
$\partial^0 \rho+\partial^i(\rho u^i)=0$ in Eq.\ \eqref{micro} 
and comparing to Eq.\ \eqref{Ta_nonrel},
we identify  $m\langle{\frac\hbar2\ms^i}\rangle=\rho\, \ell^i$,
$C^{ji}=-\langle{\frac\hbar2\ms^ip^j}\rangle+m\langle \frac\hbar2 \ms^i\rangle u^j$, and
$T^{jk}=-T_{\mathrm{HW}}^{jk}$.

\textit{Conclusions} --
In this letter, we computed
the collision term in the Boltzmann equation up to first order in $\hbar$,
accounting for the nonlocality of the microscopic collision process.
Nonlocal collisions are essential to convert
orbital into spin angular momentum.
We have shown that the collision term vanishes in global 
equilibrium, where the spin potential is equal to the thermal vorticity. 
In the approach to equilibrium, a rotating fluid of particles will develop a
nonvanishing polarization, while a polarized fluid will develop a nonvanishing vorticity.
Furthermore, we have shown that, in a certain pseudo-gauge
choice~\cite{hilgevoord1965covariant}, the antisymmetric part of
the energy-momentum tensor arises solely from the nonlocal contribution 
to the collision term. 
The equations of motion for the energy-momentum and spin tensors show that, 
away from global equilibrium, nonlocal collisions always imply dissipative dynamics.
In the nonrelativistic limit, we have obtained well-known results for 
hydrodynamics with internal degrees
of freedom, as has been applied to e.g.\ micropolar fluids~\cite{Lukaszewicz1999},
spintronics~\cite{takahashi2016spin}, and chiral active fluids~\cite{banerjee2017odd}.
An interesting extension of our work would be to 
derive the equations of motion of relativistic dissipative spin hydrodynamics 
using the method of
moments~\cite{Denicol:2012cn}.

\textit{Acknowledgments} --
The authors thank F.\ Becattini, W.\ Florkowski, X.\ Guo, U.\ Heinz,
Y.-C.\ Liu, R.\ Ryblewski, L.\ Tinti, and J.-J. Zhang, for enlightening discussions.
The work of D.H.R., E.S., and N.W.\ is supported by the
Deutsche Forschungsgemeinschaft (DFG, German Research Foundation)
through the Collaborative Research Center CRC-TR 211 ``Strong-interaction matter
under extreme conditions'' -- project number 315477589 - TRR 211.
E.S.\ acknowledges support by
BMBF ``Forschungsprojekt: 05P2018 - Ausbau von ALICE am LHC (05P18RFCA1)".
X.-L.S. and Q.W. are supported in part by the National Natural Science Foundation of China (NSFC) under Grant No.\ 11890713 (a subgrant of 11890710), and 11947301.

\bibliography{biblio_paper}{}

\end{document}